\documentstyle[twoside,fleqn,npb,epsfig,axodraw]{article}
\title{NNLO Corrections to the Higgs Production Cross Section.}

\author{V. Ravindran,\address{Harish-Chandra Research Institute, Chhatnag Road, 
Jhusii, Allahabad, 211019, India.}%
        \thanks{Deutsches Elektronen-Synchroton DESY, Platanenallee 6,
15738 Zeuthen.}
        J. Smith,\address{C.N. Yang Institute for Theoretical Physics,
State University of New York at Stony Brook, New York 11794-3840, USA}%
        W.L. van Neerven.\address{Instituut-Lorentz, Universiteit Leiden, 
PO Box 9506, 2300 RA Leiden, The Netherlands.}}

\begin{document}

\begin{abstract}
We discuss the next-to-next-to-leading order (NNLO) corrections to
the total cross section for (pseudo-) scalar Higgs boson production.
The computation is carried out in the effective Lagrangian approach
which emerges from the standard model by taking the limit
$m_t \rightarrow \infty$ where $m_t$ denotes the mass of the top quark.

\end{abstract}

\maketitle


The Higgs boson (${\rm H}$) has not yet been discovered.  Recently a new central 
value for the top quark mass of $m_t=178.0~{\rm GeV/c^2}$ was 
reported \cite{top} so that the lower limit on the mass of the Higgs from the 
LEP experiments is now $m\sim 117~{\rm GeV/c^2}$ \cite{new}. 
The search for the Higgs boson is a high priority item at the Tevatron and at
the future LHC. 

In this contribution we discuss H production via the gluon-gluon fusion
mechanism. In the standard model the H-g-g coupling contains 
quark loops.  Since the ${\rm H-q-\bar q}$ coupling 
is proportional to the mass of the quark the top-quark loop gives 
the largest cross section.  The calculations simplify if one takes the
limit $m_t \rightarrow \infty$ and uses the heavy quark effective 
theory (HQET) containing a direct Hgg coupling. Studies show that this   
is also an excellent approximation.
The Lagrangian density is
\begin{eqnarray}
\label{eqn2.1}
&& {\cal L}^{\rm H}_{eff}=G_{\rm H}\,\Phi^{\rm H}(x)\,O(x)\,, 
\end{eqnarray}
with 
\begin{eqnarray}
\label{eqn2.2}
\quad O(x)=-\frac{1}{4}\,G_{\mu\nu}^a(x)\,G^{a,\mu\nu}(x)\,,
\end{eqnarray}
whereas pseudo-scalar Higgs (${\rm A}$) production is obtained from
\begin{eqnarray}
\label{eqn2.3}
&&{\cal L}_{eff}^{\rm A}=\Phi^{\rm A}(x)\Bigg [G_{\rm A}\,O_1(x)+
\tilde G_{\rm A}\,O_2(x)\Bigg ]  
\end{eqnarray}
with
\begin{eqnarray}
\label{eqn2.4}
&&O_1(x)=-\frac{1}{8}\,\epsilon_{\mu\nu\lambda\sigma}\,G_a^{\mu\nu}\,
G_a^{\lambda\sigma}(x) \,,
\nonumber\\[2ex]
&&O_2(x) =-\frac{1}{2}\,\partial^{\mu}\,\sum_{i=1}^{n_f}
\bar q_i(x)\,\gamma_{\mu}\,\gamma_5\,q_i(x)\,,
\end{eqnarray}
where $\Phi^{\rm H}(x)$ and  $\Phi^{\rm A}(x)$ represent the scalar and 
pseudo-scalar fields respectively and $n_f$ denotes the number of light 
flavours.
Furthermore the gluon field strength is given by $G_a^{\mu\nu}$ and the 
quark field is denoted by $q_i$.
The factors multiplying the operators are chosen in such a way that the 
vertices are normalised to the effective coupling constants $G_{\rm H}$, 
$G_{\rm A}$ and $\tilde G_{\rm A}$. The latter are determined by the 
top-quark triangular loop graph, including all QCD corrections, taken in the
limit $m_t\rightarrow \infty$ which describes the decay process 
${\rm B} \rightarrow g + g$ with ${\rm B}={\rm H},{\rm A}$ namely
\begin{eqnarray}
\label{eqn2.5}
&&G_{\rm B}=-2^{5/4}\,a_s(\mu_r^2)\,G_F^{1/2}\,
\tau_{\rm B}\,F_{\rm B}(\tau_{\rm B})\,
\nonumber\\[2ex]&&
\times {\cal C}_{\rm B}
\left (a_s(\mu_r^2),\frac{\mu_r^2}{m_t^2}\right )\quad ,\quad
\tau=\frac{4m_t^2}{m^2}\,,
\nonumber\\[2ex]
&&\frac{\tilde G_{\rm A}}{G_{\rm A}}
=-\Bigg [a_s(\mu_r^2)C_F\left (\frac{3}{2}-3
\ln \frac{\mu_r^2}{m_t^2}\right )+.. \Bigg ]
\end{eqnarray}
where $a_s(\mu_r^2)=\alpha_s(\mu_r^2)/4\pi$ and $\mu_r$ denotes 
the renormalization scale. Further $G_F$ represents the Fermi constant and the
functions $F_{\rm B}$ are given in the asymptotic limit by
\begin{eqnarray}
\label{eqn2.6}
 \mathop{\mbox{lim}}\limits_{\vphantom{\frac{A}{A}} \tau \rightarrow \infty}
F_{\rm H}(\tau)=\frac{2}{3\,\tau}\,, 
 \mathop{\mbox{lim}}\limits_{\vphantom{\frac{A}{A}} \tau \rightarrow \infty}
F_{\rm A}(\tau)=\frac{1}{\tau}\,\cot \beta\,. 
\end{eqnarray}
$\cot \beta$ denotes the mixing angle in the Two-Higgs-Doublet Model
\cite{2hdm}.
The coefficient functions ${\cal C}_{\rm B}$ originate from the corrections 
to the top-quark triangular graph provided one takes the 
limit $m_t\rightarrow \infty$.  They were computed
up to order $\alpha_s^2$ in \cite{kls}, \cite{cks} for the H 
and in \cite{cksb} for the A. 

\begin{figure}
\begin{center}
\begin{picture}(130,130)(0,0)
\ArrowLine(0,110)(28,110)
\ArrowLine(120,20)(92,20)
\CArc(45,110)(17,120,240)
\CArc(25,110)(17,300,60)

\CArc(95,20)(17,120,240)
\CArc(75,20)(17,300,60)

\Gluon(40,100)(60,65){3}{5}
\Gluon(80,30)(60,65){3}{5}

\DashArrowLine(42,115)(110,115){3}
\DashArrowLine(42,110)(110,110){3}
\DashArrowLine(42,105)(110,105){3}

\DashArrowLine(78,25)(10,25){3}
\DashArrowLine(78,20)(10,20){3}
\DashArrowLine(78,15)(10,15){3}

\DashArrowLine(60,65)(90,65){3}
\Gluon(48,90)(80,90){3}{5}
\Gluon(67,50)(35,50){3}{5}

\Text(0,120)[t]{$p_1(P_1)$}
\Text(130,20)[t]{$p_2(P_2)$}
\Text(40,85)[t]{$g$}
\Text(80,50)[t]{$g$}
\Text(110,68)[t]{$H$}

\Text(85,95)[t]{$g$}
\Text(30,55)[t]{$g$}

\end{picture}
\caption[]{The reaction $p_1(P_1) + p_2(P_2) \rightarrow H  + `X'$. }
\label{fig1}
\end{center}
\end{figure}
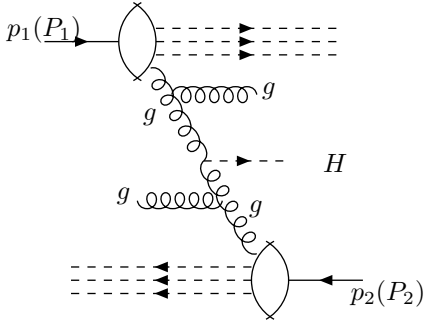

One tree diagram contributing to the reaction 
$p_1(P_1)+p_2(P_2) \rightarrow H + 'X'$ 
is shown in Fig. 1. 
On the Born level we have the partonic reaction
\begin{eqnarray}
\label{eqn2.7}
g(p_1)+g(p_2) \rightarrow {\rm B}(p_5)\,.
\end{eqnarray}
In NLO we have in addition to the one-loop virtual corrections to the above
reaction the following two-to-two body processes
\begin{eqnarray}
\label{eqn2.8}
&&g+g \rightarrow {\rm B} + g \quad\,,\quad
g+q(\bar q) \rightarrow {\rm B} + q(\bar q)\,,
\nonumber\\[2ex]
&&\quad q+ \bar q \rightarrow {\rm B} + g\,.
\end{eqnarray}
Such diagrams contain the same basic
building blocks as in the Drell-Yan production of dileptons
(see the contribution of van Neerven to these proceedings).
The LO and NLO contributions to the total cross section
for H production were computed in \cite{dawson}. 
The two-to-three body processes were computed 
for H and A production in 
\cite{kdr} and \cite{kade} respectively using helicity methods. 
The one-loop corrections to the two-to-two body reactions above 
were computed for the H in \cite{schmidt} and the A in \cite{fism}. 
These matrix elements were used to compute the transverse 
momentum and rapidity distributions of the H up to NLO in
\cite{fgk}, \cite{rasm1} and the A in \cite{fism}.

In NNLO we receive contributions from the two-loop virtual corrections to
the Born process in Eq. (\ref{eqn2.7}) and the one-loop corrections to
the reactions in  Eq. (\ref{eqn2.8}). To these contributions one has to add
the results obtained from several two-to-three body reactions containing
gluons and quarks, e.g.  
\begin{eqnarray}
\label{eqn2.9}
g(p_1) + g(p_2) \rightarrow g(p_3) + g(p_4)+{\rm B}(p_5)\,.
\end{eqnarray} 
The effective Lagrangian method was also applied to obtain the NNLO total 
cross section for H production by the calculation of the two-loop 
corrections to the H-g-g vertex in \cite{harland}, the soft-plus-virtual gluon 
corrections in \cite{cafl1}, \cite{haki1} and the computation of the 
two-to-three 
body processes in \cite{haki2}, \cite{anme1}, \cite{rsvn0}, \cite{rsvn1}. 
These calculations were 
repeated for pseudo-scalar Higgs production in \cite{haki3}, \cite{anme2}
and \cite{rsvn0}, \cite{rsvn1}.
In the case of A production one also has to add the
contributions due to interference terms coming from the operators $O_1$ 
and $O_2$ in Eq. (4), (see Fig. 1b and Fig. 2b of \cite {haki3}). 

Let us start with the NNLO virtual corrections.
The most complicated diagram is the two-loop non-planar graph for 
$H(p_2-p_1)\rightarrow g(p_1) + g(-p_2)$ shown in Fig.2. The momenta
are incoming and $k_1$, $k_2$ label the two loop momenta.

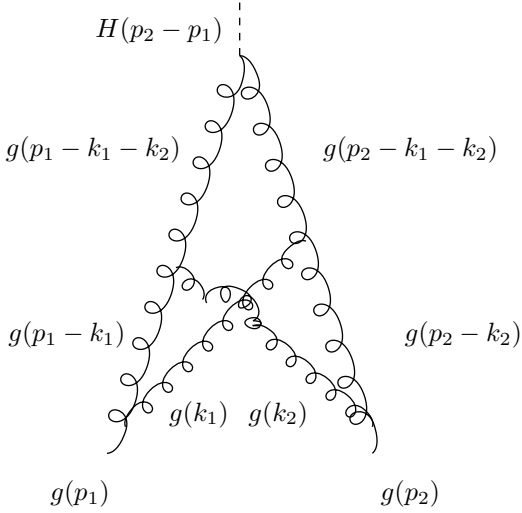
\begin{figure}
\begin{center}
\begin{picture}(230,180)(0,0)

\DashLine(85,200)(85,180){3}

\Gluon(85,180)(35,30){4}{10}
\Gluon(85,180)(135,30){4}{10}

\Gluon(61,100)(71,88){3}{1}
\Gluon(90,78)(135,40){3}{5}

\Gluon(42,40)(110,110){3}{8}

\GlueArc(80,80)(10,0,138){3}{2}
\GlueArc(80,80)(10,350,360){3}{1}

\Text(55,195)[t]{$H(p_2-p_1)$}

\Text(30,150)[t]{$g(p_1-k_1-k_2)$}
\Text(150,150)[t]{$g(p_2-k_1-k_2)$}

\Text(20,80)[t]{$g(p_1-k_1)$}
\Text(170,80)[t]{$g(p_2-k_2)$}

\Text(70,50)[t]{$g(k_1)$}
\Text(100,50)[t]{$g(k_2)$}

\Text(25,20)[t]{$g(p_1)$}
\Text(150,20)[t]{$g(p_2)$}

\end{picture}
\caption[]{The two-loop non-planar crossed graph. }
\label{fig2}
\end{center}
\end{figure}

The basic integral for this diagram has been known for a long time. It is
\begin{eqnarray}
\label{eqn2.10}
&&   V_{123567} =
       s^{-2}\,s^\varepsilon\,S_n^2 \,
\bigg[  - 16\,\varepsilon^{-4} + 24\,\varepsilon^{-2}\,\zeta(2)
\nonumber\\[2ex]&&
- \frac{166}{3}\,\varepsilon^{-1}\,\zeta(3)
+ \frac{177}{10}\,\zeta(2)^2 \bigg]\,.
\end{eqnarray}
where $s=(p_1-p_2)^2$ and 
$S_n$ is the angular factor in $n=4+\varepsilon$ dimensions.
In Drell-Yan production of lepton pairs the highest power of momenta in
the numerator was three which required the result 
\begin{eqnarray}
\label{eqn2.11}
&&   (k_1.p_2)^3\,V_{123567} =
       s\, s^\varepsilon\,S_n^2 \,
\bigg[  \frac{69}{8} +\varepsilon^{-4} 
- \frac{3}{2}\,\varepsilon^{-3}
\nonumber\\[2ex]&&
-\frac{3}{2}\,\varepsilon^{-2}\,\zeta(2) + 3\,\varepsilon^{-2}
+ \frac{15}{8}\,\varepsilon^{-1}\,\zeta(2)
\nonumber\\[2ex]&&
+ \frac{83}{24}\,\varepsilon^{-1}\,\zeta(3)
- \frac{21}{4}\,\varepsilon^{-1} 
- 3\,\zeta(2)
\nonumber\\[2ex]&&
- \frac{177}{160}\,\zeta(2)^2 -5\,\zeta(3) \bigg]\,.
\end{eqnarray}
However in the Higgs case we need the integrals with up to four powers in the
numerator such as 
\begin{eqnarray}
\label{eqn2.12}
&&   (k_1.p_2)^4V_{123567} =
        s^2\,s^\varepsilon\,S_n^2 \, \bigg[  - \frac{56087}{10368}
- \frac{1}{2}\,\varepsilon^{-4} 
\nonumber\\[2ex]&&
+ \frac{11}{12}\,\varepsilon^{-3}
+ \frac{3}{4}\,\varepsilon^{-2}\,\zeta(2) - \frac{67}{36}\,\varepsilon^{-2}
- \frac{55}{48}\,\varepsilon^{-1}\,\zeta(2)
\nonumber\\[2ex]&&
- \frac{83}{48}\,\varepsilon^{-1}\,\zeta(3)
+ \frac{2837}{864}\,\varepsilon^{-1} + \frac{67}{36}\,\zeta(2)
\nonumber\\[2ex]&&
+ \frac{177}{320}\,\zeta(2)^2 + \frac{55}{18}\,\zeta(3) \bigg]\,.
\end{eqnarray}

The virtual corrections were calculated in \cite{harland}
with the program Mincer \cite{mincer} 
and we have checked them by using the methods in \cite{vn1}.
In the pseudoscalar case everyone used
the HVBM prescription for the $\gamma_5$-matrix and the Levi-Civita tensor in 
\cite{hv}, \cite{brma}, \cite{akde}.  

The evaluation of the two-to-three body phase space integrals
is possible in suitable frames.
Since we integrate over the total phase space the integrals are Lorentz
invariant and therefore frame independent. The matrix elements of the partonic
reactions can be partial fractioned in such a way that maximally two
factors $P_{ij}=(p_i+p_j)^2$ depend on the 
polar angle $\theta$ and the azimuthal angle $\phi$. 
Furthermore only one factor depends on
$\theta$ whereas the other one depends both on $\theta$ and $\phi$.
Therefore the following combinations show up in the matrix elements
\begin{eqnarray}
\label{eqn2.13}
&&P_{ij}^k\,P_{mn}^l \quad , \quad 
P_{ij}^k\,P_{m5}^l \quad , \quad P_{i5}^k\,P_{m5}^l\,,
\nonumber\\[2ex]
&& 4\ge k\ge -2 \quad , \quad 4\ge l\ge -2\,, 
\nonumber\\[2ex]
&&p_i^2=p_j^2=p_m^2=p_n^2=0 \quad\,,\quad p_5^2=m^2\,.
\end{eqnarray}
For the first combination it is easy to perform the angular integrations
exactly in the CM frame of the incoming partons
since all momenta represent massless particles. This requires the result 
in \cite{vn1}.
\begin{eqnarray}
\label{eqn2.14}
&&\int_0^\pi d \theta \int_0^\pi d\phi 
\sin^{n-3}\theta \sin^{n-4}\phi  {\cal C}_{kl}(\theta,\phi,\chi)=
\nonumber\\ [2ex] && 
2^{1-k-l} \pi  
{\Gamma({1 \over 2}n-1-k)\Gamma({1 \over 2}n-1-l)
\Gamma(n-3) \over \Gamma(n-2-k-l)  
\Gamma^2({1 \over 2}n-1) }
\nonumber\\ [2ex] &&
\times F_{1,2}\Big(k,l,{1\over 2} n-1;\cos^2(\chi/2)\Big)\,,
\end{eqnarray}
where the function
${\cal C}_{ij}(\theta,\phi,\chi) =(1-\cos\theta)^{-k}$ 
$ (1-\cos\theta\cos\chi -\sin\theta\cos\phi\sin\chi)^{-l}$. 
Here $\cos\chi$ is 
related to kinematical variables such as $x~(=m^2/s$, $\sqrt(s)$=CM energy) 
and another two integration variables $z$ and $y$, and $F_{1,2}(a,b,c;t)$ 
is the hypergeometric function. 

\begin{figure}[htb]
\vspace{9pt}
\centering{\rotatebox{270}{\epsfig{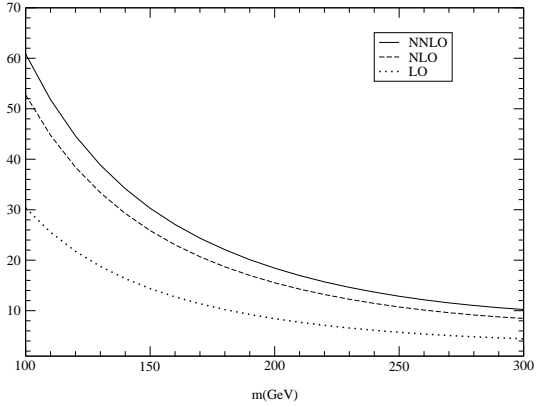}}}
\caption{The total cross section $\sigma_{\rm tot}$ with all channels
included plotted as a function of the Higgs mass at $\sqrt{S}=14$ TeV
with $\mu=m$}
\label{fig:fig3}
\end{figure}

The angular integral of the second combination is more difficult to compute
because one particle is massive and the result is an one dimensional integral 
over a hypergeometric function
which however can be expanded around $\varepsilon$. Examples of these types
of integrals can be found in Appendix C of \cite{bkns}. 
The last combination is very difficult to compute in n dimensions 
because both factors contain the massive particle indicated by $p_5$. 
This combination can be avoided if 
one chooses one of the following three frames, namely 
(1) the CM frame of the incoming partons, 
(2) the CM frame of the two outgoing
partons indicated by the momenta $p_3$ and $p_4$,
and (3) the CM frame of one of the outgoing
partons and the H indicated by the momenta $p_4$ and 
$p_5$ respectively

\begin{figure}[htb]
\vspace{9pt}
\centering{\rotatebox{270}{\epsfig{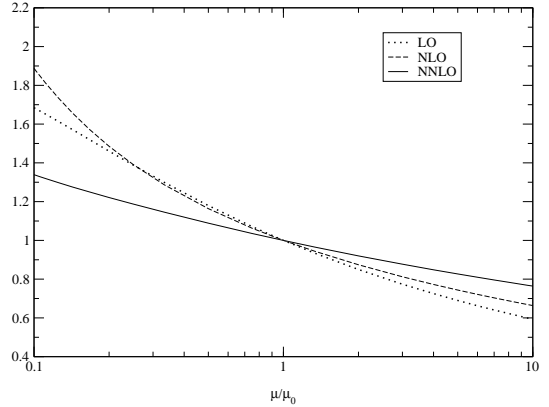}}}
\caption{The quantity $N(\mu/\mu_0)$ plotted in the range
$0.1 < \mu/\mu_0 < 10$ with $\mu_0=m$ and $m=100$ ${\rm GeV/c}^2$.}  
\label{fig:fig4}
\end{figure}
  
In (3) we performed one angular integration (say $\phi$)
exactly and performed the remaining $\theta$ integration after 
expanding the integrands in powers of $\varepsilon=n-4$. 
In all these frames, using various Kummer's relations,
the hypergeometric functions are simplified to the form 
$F_{1,2}(\pm\varepsilon/2,\pm\varepsilon/2,1\pm\varepsilon/2;f(x,y,z))$
which is the most suitable for integrations over $z$ and $y$.
The higher powers of $1/(1-z)^{\alpha+\beta\varepsilon}$
or $1/z^{\alpha+\beta\varepsilon}$, where $\alpha > 1$, were then reduced 
by successive integration by parts with exact 
hypergeometric functions until we arrived at terms in
$1/(1-z)^{1+\beta \varepsilon}$ or
$1/z^{1+\beta\varepsilon}$ multiplied by regular functions.
This required the following identity 
\begin{eqnarray}
\label{eqn2.15}
&& {d \over dz}F_{1,2}(\varepsilon/2,\varepsilon/2,1+\varepsilon/2; f(z)) =
\nonumber\\ &&
 {\varepsilon \over 2 f(z)} {d f(z)\over dz}\Big(\big(1
 -f(z)\big)^{-{\varepsilon \over 2}}
\nonumber\\ &&
-F_{1,2}\big(\varepsilon/2,\varepsilon/2,1+\varepsilon/2;f(z)\big)\Big)
\,.
\end{eqnarray}
The remaining integrals like $\int_0^1 dz z^{-1-\beta \varepsilon}f(z)$ 
and/or $\int_0^1 dz (1-z)^{-1-\beta \varepsilon}f(z)$ 
were simplied as follows:
\begin{eqnarray}
\label{eqn2.16}
&&\int_0^1 dz  z^{-1-\beta \varepsilon}
f(z) = 
\nonumber\\ &&
 \int_0^\delta dz ~z^{-1-\beta \varepsilon} f(z)
 + \int_\delta^1 dz~ z^{-1-\beta \varepsilon} f(z) \,,
\end{eqnarray}
where $\delta << 1$.
The first term can be evaluated to be
$f(0) ( [-\beta \varepsilon]^{-1}+\ln\delta+
[-\beta \varepsilon / 2] \ln^2\delta+\cdot \cdot \cdot)$.
After expanding $z^{-\beta \varepsilon} $ in powers
of $\varepsilon$ in the second term the $z$ integration can be
performed exactly order-by-order in $\varepsilon$ with non-zero $\delta$.
At the end the $\delta$ dependence cancels in each order in $\varepsilon$.
Since the $z$ integration over the hypergeometric functions
is nontrivial due to their complicated arguments, we expanded them in
powers of $\varepsilon$ prior to the $z$ integration, using
\begin{eqnarray}
\label{eqn2.17}
&&F_{1,2}( \varepsilon/2, \varepsilon/2,1+ {\varepsilon \over 2};
f(z))= 1
\nonumber \\&&
+ {\varepsilon^2 \over 4} {\rm Li}_2(f(z))
+ {\varepsilon^3 \over 8} \Big({\rm S}_{1,2}(f(z))
- {\rm Li}_{3}(f(z))\Big)
\nonumber \\ &&
+{\varepsilon^4 \over 16} \Big({\rm S}_{1,3}(f(z))-{\rm S}_{2,2}
(f(z)) +{\rm Li}_4(f(z))\Big) 
\nonumber \\ && \cdots \,,
\end{eqnarray}
where S$_{n,p}(z)$ is the Nielsen function,
with $ n,p \geq 1$ and
${\rm Li}_n(z)={\rm S}_{n-1,1}(z)$(see \cite{lewin}).

We have repeated the same procedure to perform the remaining $y$ integration.
The integrals were programmed using FORM \cite{form} by extending
a program originally used to compute the NNLO coefficient functions of 
the Drell-Yan process in \cite{mmn1}, \cite{ham}. 
After performing all these integrals,
we removed all the ultraviolet divergences by coupling constant
and operator renormalisation.  The remaining collinear divergences
are removed by mass factorisation. Then we are left with finite
partonic cross sections which are folded with parton distribution functions
to the compute hadronic cross section for inclusive H and A production.

Our method differs from the one presented in \cite{haki2}, \cite{haki3} 
and the approach followed in \cite{anme1}, \cite{anme2}. The authors in 
the latter references compute the total
cross section using the Cutkosky \cite{cut} technique where one- and 
two-loop Feynman integrals are cut in certain ways. 
These Feynman integrals can be computed using various techniques (for more
details see the references in \cite{anme1}).  The authors
in \cite{haki2} chose the more conventional method which was already used  
to compute the coefficient
functions for the Drell-Yan process. However instead of an exact
computation of the $2\rightarrow 2$ and $2\rightarrow 3$ body phase space 
integrals they expanded them around $x=m^2/s=1$. 
Since the coefficient functions were known from the Drell-Yan reaction
to be expressible in a finite number of polylogarithms like  
${\rm Li}_n(x)$, ${\rm S}_{n,p}(x)$  and logarithms of 
the types $\ln^i x~\ln^j(1-x)$, which are all multiplied by polynomials in $x$,
one could expand these functions in the limit 
$x\rightarrow 1$ and match them with the expressions coming from the
phase space integrals. In this way the coefficients in the expansion
were determined. 

\begin{figure}[htb]
\vspace{9pt}
\centering{\rotatebox{270}{\epsfig{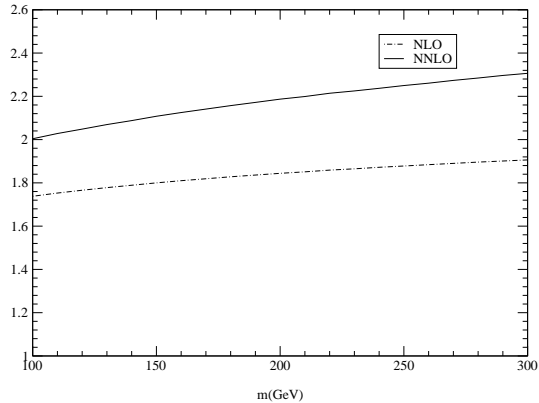}}}
\caption{The $K$-factors in NLO and NNLO at $\sqrt{S}=14 {\rm TeV}$ as a
function of the Higgs mass using the MRST parton density sets;
$K^{\rm NNLO}$ solid line, $K^{\rm NLO}$ dot-dashed line.}
\label{fig:fig5}
\end{figure}

In Fig.3 we show the total cross section for H production in 
p-p collisions at the LHC.
For the computation of the effective coupling constant $G_{\rm B}$ in Eq. 
(\ref{eqn2.3}) we choose the top quark mass $m_t=173.4~{\rm GeV/c^2}$ and the
Fermi constant $G_F=1.16639\times 10^{-5}~{\rm GeV}^{-2}=4541.68~{\rm pb}$. 
We used the parton densities in \cite{mrst1}.
The cross sections for the A are about $9/4~\cot^2 \beta$ times 
those for the H.
Next we study their variation with respect to the scale $\mu$ by 
computing the ratio
\begin{eqnarray}
\label{eqn1.19}
N\left (\frac{\mu}{\mu_0}\right )=\frac{\sigma_{\rm tot}(\mu)}
{\sigma_{\rm tot}(\mu_0)}\,,
\end{eqnarray}
where $\mu_0=m$ and $\mu$ is varied in the range $0.1<\mu/\mu_0<10$.
In Fig. 4 a plot of $N$ is shown for $m=100~{\rm GeV/c^2}$. Here one observes
a clear improvement while going from LO to NNLO. In particular the curve
for NNLO is flatter than that for NLO. 
Another way to estimate the reliability of this result is to study the 
rate of convergence of the perturbation series, which is 
represented by the $K$-factor. We choose the following definitions
\begin{eqnarray}
\label{eqn1.20}
K^{\rm NLO}=\frac{\sigma_{\rm tot}^{\rm NLO}}{\sigma_{\rm tot}^{\rm LO}} 
\quad\,,\quad
K^{\rm NNLO}=\frac{\sigma_{\rm tot}^{\rm NNLO}}{\sigma_{\rm tot}^{\rm LO}}\,. 
\end{eqnarray}
In Fig. 5 one observes that both $K$-factors vary slowly as 
$m$ increases. Moreover there is considerable improvement in the rate of
convergence if one goes from NLO to NNLO. At $m=100~{\rm GeV/c^2}$
we have $K^{\rm NLO}\sim 1.74$ whereas $K^{\rm NNLO}\sim 2.00$. Still
the corrections for Higgs boson production are larger than those
obtained from $Z$-boson production at the LHC where one gets
$K^{\rm NLO}\sim 1.22$ and $K^{\rm NNLO}\sim 1.16$ (see \cite{ham}).


\begin{thebibliography}{99}
%
\bibitem{top}
Tevatron Electroweak Working Group, [D0 Collaboration], hep-ex/0404010.
\bibitem{new}
R. Barate et al, [ALEPH Collaboration], Phys. Lett. B 565 (2003) 61,
hep-ex/0306033.
\bibitem{2hdm} J.F. Gunion, H.E. Haber, G.L. Kane, S. Dawson,
"The Higgs Hunter's Guide", Addison-Wesley, Reading 1990.
\bibitem{kls}
M. Kr\"amer, E. Laenen, M. Spira, Nucl. Phys. B511 (1998) 523,
hep-ph/9611272.
\bibitem{cks}
K.G. Chetyrkin, B.A. Kniehl, M. Steinhauser, Phys. Rev. Lett. 79
(1997) 353, hep-ph/9705240.
\bibitem{cksb}
K.G. Chetyrkin, B.A. Kniehl, M. Steinhauser, W.A. Bardeen, Nucl. Phys. B535
(1998) 3, hep-ph/9807241.
\bibitem{dawson}
S. Dawson, Nucl. Phys. B359 (1991) 283;\\
A. Djouadi, M. Spira, P. Zerwas, Phys. Lett. B264 (1991) 440.
\bibitem{kdr}
R.P. Kauffman, S.V. Desai, D. Risal, Phys. Rev. D55 (1997) 4005;
Erratum-ibid, D58 (1998) 119901, hep-ph/9610541.
\bibitem{kade}
R.P. Kauffman, S.V. Desai, Phys. D59 (1999) 05704, hep-ph/9808286.
\bibitem{schmidt}
C.R. Schmidt, Phys. Lett. B413 (1997) 391, hep-ph/9707448.
\bibitem{fism}
B. Field, J. Smith, M.E. Tejeda-Yeomans, W.L. van Neerven, Phys. 
Lett. B551 (2002) 137, hep-ph/0210369.
\bibitem{fgk}
D. de Florian, M. Grazzini, Z. Kunszt, Phys. Rev. Lett. 82 (1999)
5209, hep-ph/9902483.
\bibitem{rasm1}
V. Ravindran, J. Smith, W.L. van Neerven, Nucl. Phys. B634 (2002) 247,
hep-ph/0201114.
\bibitem{harland}
R.V. Harlander, Phys. Lett. B492 (2000) 74, hep-ph/0007289.
\bibitem{cafl1}
S. Catani, D. de Florian, M. Grazzini, JHEP 0105 (2001) 025, 
hep-ph/0102227.
\bibitem{haki1}
R.V. Harlander, W.B. Kilgore, Phys. Rev. D64 (2001) 013015, 
hep-ph/0102241.
\bibitem{haki2}
R.V. Harlander, W.B. Kilgore, Phys. Rev. Lett. 88 (2002) 201801,
hep-ph/0201206.
\bibitem{anme1}
C. Anastasiou, K. Melnikov, Nucl. Phys. B646 (2002) 220, 
hep-ph/0207004.
\bibitem{rsvn0}
V. Ravindran, J. Smith, W.L. van Neerven, Nucl. Phys. B665 (2003) 325.
hep-ph/0302135.
\bibitem{rsvn1}
V. Ravindran, J. Smith, W.L. van Neerven, Pramana 62 (2004) 683.
hep-ph/0304005.
\bibitem{haki3}
R.V. Harlander, W.B. Kilgore, JHEP 0210 (2002) 017, hep-ph/0208096.
\bibitem{anme2}
C. Anastasiou, K. Melnikov, Phys. Rev. D67 (2003) 037501, hep-ph/0208115.
\bibitem{mincer}
S.G. Gorishnii et al, Comput. Phys. Commun. 55 (1989) 381;
S.A. Larin, F.V. Tkachov and J.A.M. Vermaseren,
Rep, No. NIKHEF-H/91-18 (Amsterdam, 1991); available from
http://www.nikhef.nl/~form.
\bibitem{vn1}
W. L. van Neerven, Nucl. Phys. B268 (1986) 453.
\bibitem{hv}
G. 't Hooft, M. Veltman, Nucl. Phys. B44 (1972) 189.
\bibitem{brma}
P. Breitenlohner, B. Maison, Commun. Math. 53 (1977) 11, 39, 55.
\bibitem{akde}
D. Akyeampong, R. Delbourgo, Nuov. Cim. 17A (1973) 578, 18A (1973) 94, 19A
(1974) 219.
\bibitem{bkns}
W. Beenakker, H. Kuijf, W.L. van Neerven, J. Smith, Phys. Rev. D40
(1989) 54.
\bibitem{lewin}
L. Lewin, "Polylogarithms and Associated Functions", North Holland,
Amsterdam, 1983.
\bibitem{form}
J.A.M. Vermaseren, Nucl. Phys. Proc. Suppl. 116 (2003) 343,
hep-ph/0211297; version 3.0 available from http://www.nikhef.nl/~form.
\bibitem{mmn1}
T. Matsuura, S.C. van der Marck, W.L. van Neerven, Phys. Lett. B211
(1988) 171; ibid Nucl. Phys.  B319 (1989) 570.
\bibitem{ham}
R. Hamberg, W.L. van Neerven, T. Matsuura, Nucl. Phys. B359
(1991) 343; Erratum-ibid Nucl. Phys. B644 (2002) 403.
\bibitem{cut}
R.E. Cutkosky, J. Math. Phys. 1 (1960) 429.
\bibitem{mrst1}
A.D. Martin, R.G. Roberts, W.J. Stirling, R.S. Thorne, Phys. Lett.
B531 (2002) 216, hep-ph/0201127; Eur. Phys. J.  C23 (2002) 73, hep-ph/0110215.
\end{thebibliography}
\end{document}